\documentclass[twocolumn,aps,superscriptaddress,showpacs,floatfix]{revtex4-1}
\usepackage{graphicx}
\usepackage{dcolumn}
\usepackage{bm}
\usepackage[colorlinks,linkcolor=blue,anchorcolor=blue,citecolor=blue,urlcolor=blue]{hyperref}
\usepackage{amsmath}
\usepackage{multirow}
\usepackage{longtable}
\usepackage{supertabular}
\usepackage{CJK,upgreek,fancyhdr}
\usepackage{lastpage}
\usepackage{longtable}
\usepackage{changes}
\usepackage{etoolbox}
\usepackage{setspace}
\usepackage{tabularx}
\usepackage{lineno,hyperref}
\usepackage{supertabular}
\usepackage[normalem]{ulem}
\usepackage{eqnarray}
\usepackage{booktabs} 
\usepackage{subfigure}
\usepackage{amsmath}
\usepackage{float}
\usepackage{placeins}

\begin{document}

\preprint{APS/123-QED}

\title{Inelastic electron scattering-induced nuclear excitation rates and dynamics in $^{229}$Th}

\author{Yang-Yang Xu}
\affiliation{Department of Physics, National University of Defense Technology, 410073 Changsha, People's Republic of China}
\author{Jun-Hao Cheng}
\affiliation{Department of Physics, National University of Defense Technology, 410073 Changsha, People's Republic of China}
\author{You-Tian Zou}
\affiliation{Department of Physics, National University of Defense Technology, 410073 Changsha, People's Republic of China}
\author{Qiong Xiao}
\affiliation{Department of Physics, National University of Defense Technology, 410073 Changsha, People's Republic of China}
\author{Tong-Pu Yu}
\email{tongpu@nudt.edu.cn}
\affiliation{Department of Physics, National University of Defense Technology, 410073 Changsha, People's Republic of China}

\begin{abstract}
In the present work, we investigate the excitation rates and population dynamics of $^{229}$Th nuclei induced by inelastic electron scattering, focusing on how electron energy, flux, and ionic charge state influence the excitation process of the nuclei. Using the Dirac Hartree-Fock-Slater method, we calculate cross sections for both the isomeric state (8.36 eV) and the second-excited state (29.19 keV) of $^{229}$Th over a wide range of ionic charge states and electron energies. Our results demonstrate that these factors significantly impact the nuclear excitation efficiency. The effect of indirect excitation through the second-excited state on enhancing the accumulation of nuclei in the isomeric state cannot be ignored. By applying rate equations to model the temporal evolution of nuclear populations, we show that under optimal conditions, up to 10\% of $^{229}$Th$^{4+}$ ions can be accumulated in the isomeric state. These findings provide important insights for optimizing electron-nucleus interactions, contributing to the development of $^{229}$Th-based nuclear clocks and relevant precision measurement applications.
\end{abstract}

\maketitle

\section{Introduction}

$^{229}$Th possesses a unique nuclear isomeric state (IS) with an excitation energy of approximately 8.35574(3) eV above its ground state, making it the lowest known nuclear excited state \cite{PhysRevLett.132.182501}. In neutral atoms, this IS has a very short half-life of about 7 microseconds and primarily decays via Internal Conversion (IC), where nuclear energy is transferred to atomic electrons, resulting in their excitation and subsequent emission \cite{PhysRevLett.118.042501}. In highly charged $^{229}$Th ions, with most or all electrons stripped away, the IC process is significantly suppressed, and radiative transitions ($\gamma$-ray emission) dominate, extending the IS's half-life to several tens of minutes or longer \cite{PhysRevC.92.054324,PhysRevC.100.054316,PhysRevC.101.054602}. This property has garnered significant interest for applications in nuclear optical clocks and precision nuclear spectroscopy. Accessing and manipulating these nuclear states presents new opportunities for high-precision measurements and advanced technologies.

Beyond the IS, ${}^{229}\mathrm{Th}$ has a second-excited state (SE) at approximately 29189.93(0.07) eV with a half-life of 82.2 picoseconds \cite{masuda2019x, PhysRevLett.125.142503}. The SE can decay radiatively into the IS, providing a pathway for indirect excitation \cite{PhysRevC.110.014330}. Traditional methods for populating the IS, such as the $\alpha$ decay of ${}^{233}\mathrm{U}$ or the $\beta$ decay of ${}^{229}\mathrm{Ac}$ \cite{PhysRevC.100.024315}, are difficult to control precisely due to their statistical nature, limiting their practical effectiveness. To overcome these challenges, x-ray pumping of the SE using synchrotron radiation has been experimentally demonstrated, offering a more controllable alternative that indirectly populates the IS through the SE's decay \cite{masuda2019x}. Another advancement is the direct laser excitation of the IS using vacuum ultraviolet (VUV) light in thorium-doped transparent crystals, where a VUV frequency comb was used to measure the ${}^{229}\mathrm{Th}$ nuclear clock transition \cite{zhang2024frequency}, and resonance fluorescence was observed in Th-doped CaF$_2$ crystals, paving the way for nuclear laser spectroscopy and nuclear clocks \cite{PhysRevLett.132.182501}.

Several electron-driven nuclear excitation processes have been proposed as well, including Nuclear Excitation by Electron Transition (NEET) \cite{10.1143/PTP.49.1574,Fujioka_1985,TKALYA1992209,PhysRevLett.85.1831,SAKABE20051}, Nuclear Excitation by Electron Capture (NEEC) \cite{GOLDANSKII1976393,CUE198925,PhysRevC.47.323,PhysRevC.59.2462,PhysRevA.73.012715,PhysRevLett.112.082501,chiara2018isomer,PhysRevLett.122.212501,PhysRevLett.128.242502}, and the Electronic Bridge (EB) process \cite{PhysRevA.81.042516,PhysRevLett.105.182501,Bilous_2018,PhysRevC.102.024604,PhysRevLett.124.192502,PhysRevLett.125.032501}. These methods exploit interactions between the nucleus and atomic electrons but often require stringent energy matching between the electronic and nuclear states, posing experimental challenges. In contrast, Nuclear Excitation by Inelastic Electron Scattering (NEIES) offers a promising approach without requiring precise resonance conditions \cite{PhysRevLett.124.242501,PhysRevC.106.044604,PhysRevC.106.064604,PhysRev.92.978}. In NEIES, incident electrons transfer energy directly to the nucleus during scattering events, potentially exciting it to higher energy states, including the IS. This mechanism allows more flexible control over the excitation process.

Previous studies on the NEIES of $^{229}$Th nuclei have primarily focused on specific energy ranges or particular excitation pathways, such as direct excitation to the IS \cite{PhysRevC.106.044604,PhysRevC.106.064604}. However, electron-induced nuclear inelastic scattering often involves multiple competing pathways, with their relative contributions dependent on electron properties. By adjusting parameters of electron, the distribution of these excitation channels can be optimized to enhance efficiency and achieve precise control over nuclear excitation. The present study analyzes the NEIES process across a broader energy range and multiple charge states, with a particular emphasis on the influence of the SE. Utilizing the Dirac-Hartree-Fock-Slater (DHFS) method \cite{PhysRev.81.385,LIBERMAN1971107}, we calculate the cross sections for electron-induced excitation of both the IS and SE of $^{229}$Th. We establish rate equations to model the temporal evolution of nuclear populations and explore how electron energy, flux, and ionic charge state affect the accumulation of nuclei in the IS through both direct and indirect excitation pathways. These results deepen our understanding of electron-nucleus interactions and help refine excitation methods. Additionally, this study contributes to future NEEC investigations by offering a clearer view of how to minimize NEIES interference, thereby improving the chances of detecting NEEC. 

This article is organized as follows. In the next section, we describe the theoretical framework for calculating the inelastic electron scattering cross sections and excitation rates in detail. In Section \ref{section 3}, the detailed calculations and discussions are provided. In Section \ref{section 4}, we provide a brief summary.

\section{THEORETICAL FRAMEWORK}

To quantitatively describe the excitation of the $^{229}$Th nucleus to its IS via electron scattering, a scattering theory framework is created and atomic units (a.u.) are used to simplify calculations, where fundamental constants like the reduced Planck constant ($\hbar$), electron mass ($m_e$), and elementary charge ($e$) equal to unity, with lengths measured in Bohr radii and energies in Hartrees. The speed of light is $c = 1/\alpha = 137.036$, where $\alpha$ is the fine-structure constant. We denote initial and final states with subscripts $i$ and $f$, while $n$ and $e$ represent nuclear and electronic quantities, respectively. The quantum numbers for the electron's total, orbital, and magnetic angular momentum are labeled as $j$, $l$, and $m$. These conventions help systematically describe the interactions and calculate the cross sections.

The differential cross section, which measures the probability of energy transfer from the incident electron to the nucleus, is given by
\begin{equation}
\label{eq1}
\frac{d \sigma}{d \Omega} = \frac{2 \pi}{v_i} \rho(E_f) \left|\langle f | H_{\mathrm{int}} | i \rangle\right|^2,
\end{equation}
where $v_i$ represents the asymptotic incoming speed of the electron, and $\rho\left(E_f\right)$ is the density of final states. The interaction Hamiltonians $H_{\text {int}}$ can be written as
\begin{equation}
\label{eq2}
\begin{aligned}
H_{\mathrm{int}}= & -\frac{1}{c} \int\left[\boldsymbol{j}_n(\boldsymbol{r})+\boldsymbol{j}_e(\boldsymbol{r})\right] \cdot \mathbf{A}(\boldsymbol{r}) d \tau \\
& +\int \frac{\rho_n(\boldsymbol{r}) \rho_e\left(\boldsymbol{r}^{\prime}\right)}{\left|\boldsymbol{r}-\boldsymbol{r}^{\prime}\right|} d \tau d \tau^{\prime}.
\end{aligned}
\end{equation}
This accounts for the coupling between the nuclear and electronic currents mediated by the radiation field's vector potential $\mathbf{A}(\boldsymbol{r})$ and the Coulomb interaction between the nucleus and the scattering electron. Here, $\boldsymbol{j}(\boldsymbol{r})$ and $\rho(\boldsymbol{r})$ represent the current densities and charge densities, respectively.

The matrix element $\langle f| H_{\text {int}}|i\rangle$ encapsulates the interaction between the incident electron and the nucleus, expanded as \cite{RevModPhys.30.353}
\begin{equation} 
\label{eq3} 
\begin{aligned} 
&\langle f| H_{\text {int }}|i\rangle= \sum_{\lambda \mu} \frac{4 \pi}{2 \lambda+1}(-1)^\mu \\
& \times\left\{\left\langle\phi_f\right| \mathcal{N}(E \lambda, \mu)\left|\phi_i\right\rangle\left\langle I_f M_f\right| \mathcal{M}(E \lambda,-\mu)\left|I_i M_i\right\rangle\right. \\
& \left.-\left\langle\phi_f\right| \mathcal{N}(M \lambda, \mu)\left|\phi_i\right\rangle\left\langle I_f M_f\right| \mathcal{M}(M \lambda,-\mu)\left|I_i M_i\right\rangle\right\}.
\end{aligned}
\end{equation}
In this expression, $\mathcal{M}$ and $\mathcal{N}$ are the electric $(E)$ and magnetic $(M)$ multipole transition operators for the nucleus and the electron, respectively. $\left|I M\right\rangle$ denotes the nuclear states with total angular momentum $I$ and its projection $M$, while $\left|\phi\right\rangle$ represents the electronic states.

With the reduced nuclear transition probabilities $B\left(E(M) \lambda ; I_i \rightarrow I_f\right)$ introduced, the differential cross section becomes
\begin{equation}
\label{eq4}
\begin{aligned}
\frac{d \sigma}{d \Omega}= & \frac{4E_i E_f}{c^4} \frac{p_f}{p_i}\times \sum_{\lambda E(M) \mu}\left\{ \frac{B\left(E(M) \lambda ; I_i \rightarrow I_f\right)}{(2 \lambda + 1)^3} \right. \\ & \left. \times \frac{1}{2} \sum_{v_i v_f} \left| \left\langle \phi_f \middle| \mathcal{N}(E(M) \lambda, \mu) \middle| \phi_i \right\rangle \right|^2 \right\},
\end{aligned}
\end{equation}
where $E$ is the total relativistic energy of electron, including both kinetic energy and rest mass energy, with $p$ being the corresponding monenta. 

In electron inelastic scattering, both initial and final electronic states are described as continuum states, modeled using Dirac distorted waves, which effectively incorporate relativistic effects. The wavefunction is expressed as follows
\begin{equation}
\label{eq5}
\begin{aligned}
|\phi\rangle=&4 \pi \sqrt{\frac{E+c^2}{2 E}}\times \\& \sum_{j l m}\left[\Omega_{j l m}^*(\mathbf{v}) \chi_v\right] e^{ \pm i \delta_{j l}}\binom{g_{j l}(r) \Omega_{j l m}(\hat{r})}{-i f_{j l^{\prime}}(r) \Omega_{j l m}(\hat{r})}.
\end{aligned}
\end{equation}
Here, $\mathbf{v}$ denotes the unit vector indicating the direction of the electron's motion, and $v$ represents its spin quantum number. The phase shift $\delta_{j l}$ accounts for the phase change of the wavefunction due to scattering. The functions $g(r)$ and $f(r)$ are the radial components of the Dirac spinor's upper and lower parts, while the spherical spinors $\Omega_{j l m}(\hat{r})$ form the angular part of the wavefunction.

As an important part of calculating the radial wave function \cite{SALVAT2019165}, the potential experienced by the electron can be derived through the Dirac-Hartree-Fock-Slater method, which includes contributions from the nucleus, electron cloud, and exchange interactions
\begin{equation}
\label{eq6}
\begin{aligned}
V_{\mathrm{DHFS}}(r)=V_{\mathrm{n}}(r)+V_{\mathrm{e}}(r)+V_{\mathrm{ex}}(r),
\end{aligned}
\end{equation}
where the nuclear potential $V_{\mathrm{n}}(r)$ is based on a Fermi distribution for nuclear charge density \cite{PhysRev.101.1131}
\begin{equation}
\label{eq7}
\begin{aligned}
\rho_{\mathrm{n}}(r)=\frac{\rho_0}{\exp \left[\left(r-R_{\mathrm{n}}\right) / z\right]+1},
\end{aligned}
\end{equation}
with $R_{\mathrm{n}}=1.07 A^{1 / 3}$ $\mathrm{fm}$ and $z=0.546$ $\mathrm{fm}$, where $A$ is the mass number. The nuclear potential is then given by
\begin{equation}
\label{eq8}
\begin{aligned}
V_{\text {n}}(r) =-\frac{e^2}{r} \int_0^r \rho_{\mathrm{n}}\left(r^{\prime}\right) 4 \pi r^{\prime 2} \mathrm{~d} r^{\prime}-e^2 \int_r^{\infty} \rho_{\mathrm{n}}\left(r^{\prime}\right) 4 \pi r^{\prime} \mathrm{d} r^{\prime}.
\end{aligned}
\end{equation}
Similarly, the electronic potential $V_{\mathrm{e}}(r)$ is obtained by integrating the electron density $\rho_e(r)$
\begin{equation}
\label{eq9}
\begin{aligned}
V_{\mathrm{e}}(r)=\frac{e^2}{r} \int_0^r \rho_e\left(r^{\prime}\right) 4 \pi r^{\prime 2} \mathrm{~d} r^{\prime}+e^2 \int_r^{\infty} \rho_e\left(r^{\prime}\right) 4 \pi r^{\prime} \mathrm{d} r^{\prime}.
\end{aligned}
\end{equation}
The exchange potential $V_{\text {ex }}(r)$, which accounts for the exchange interaction, is approximated using the Thomas-Fermi method \cite{PhysRev.81.385}
\begin{equation}
\label{eq10}
\begin{aligned}
V_{\mathrm{ex}}(r)= \begin{cases}C_{\mathrm{ex}} V_{\mathrm{ex}}^{(\mathrm{TF})}(r), &  r<r_{\text {Latter }} \\ (N+1-Z)\frac{e^2}{r}-V_{\mathrm{n}}(r)-V_{\mathrm{e}}(r), &  r \geq r_{\text {Latter }}\end{cases},
\end{aligned}
\end{equation}
where $C_{\mathrm{ex}}$=1.5 \cite{PhysRev.99.510} and $V_{\mathrm{ex}}^{(\mathrm{TF})}(r)$ is the exchange potential for a free-electron gas with $N$ and $Z$ being the number of electrons and nuclear charges, respectively. The cutoff radius $r_{\text{Latter}}$ is the outer root of the second equation in Eq.(\ref{eq10}).

Combining the relevant components, the cross section for NEIES can be expressed as
\begin{equation}
\label{eq11}
\begin{aligned}
\sigma_{\mathrm{E(M) \lambda}} = &\frac{8\pi^2}{c^4}\frac{E_i + m_e c^2}{p_i^3}\frac{E_f + m_e c^2}{p_f} \\& \times \sum_{l_i, j_i, l_f, j_f} \frac{\kappa^{2 \lambda+2}}{(2 \lambda-1)!!^2} B\left(E(M) \lambda, I_i \rightarrow I_f\right) \\
& \times \frac{\left(2 l_i+1\right)\left(2 l_f+1\right)\left(2 j_i+1\right)\left(2 j_f+1\right)}{(2 \lambda+1)^2} \\
& \times\left(\begin{array}{ccc}
	l_f & l_i & \lambda \\
	0 & 0 & 0
\end{array}\right)^2\left\{\begin{array}{ccc}
	l_i & \lambda & l_f \\
	j_f & 1 / 2 & j_i
\end{array}\right\}^2\left|T_{f i}^{E(M)\lambda}\right|^2,
\end{aligned}
\end{equation}
with the radial matrix elements being
\begin{equation}
\label{eq12}
\begin{aligned}
&T_{f i}^{E \lambda}= \int_0^{\infty} h_\lambda^{(1)}(\kappa r)\left[g_i(r) g_f(r)+f_i(r) f_f(r)\right] r^2 d r\\
& -\int_0^{\infty} h_{\lambda-1}^{(1)}(\kappa r)\frac{\kappa}{\lambda}\left[g_i(r) g_f(r)+f_i(r) f_f(r)\right] r^3 d r,
\end{aligned}
\end{equation}
and
\begin{equation}
\label{eq13}
\begin{aligned}
&T_{f i}^{M \lambda}=\frac{\eta_i+\eta_f}{\lambda} \\&\times \int_0^{\infty} h_\lambda^{(1)}(\kappa r) \left[g_i(r) f_f(r)+g_f(r) f_i(r)\right] r^2 d r.
\end{aligned}
\end{equation}
Here, $h_\lambda^{(1)}(\kappa r)$ and $h_{\lambda-1}^{(1)}(\kappa r)$ represent the spherical Hankel functions of the first kind, which describe the radial dependence of the electromagnetic field. The parameter $\eta$ is defined as $\eta=(l-j)(2j+1)$. When $\eta<0$, or in the case of magnetic multipole ($M \lambda$) transitions, the orbital angular momentum $l$ must be replaced by $l^{\prime}=2j-l$.

The NEIES excitation rate per target particle is obtained by integrating the cross section over the electron's kinetic energy, weighted by the electron flux $\Phi_e(E)$
\begin{equation}
\label{eq14}
\begin{aligned}
\lambda_{\mathrm{NEIES}}=\int dE\left[\sigma_{\mathrm{E\lambda}}(E)+\sigma_{\mathrm{M\lambda}}(E)\right] \Phi_e(E),
\end{aligned}
\end{equation}
and the total excitation rate (number of excitations per unit time) can be calculated by
\begin{equation}
\label{eq15}
\begin{aligned}
N_{\text {NEIES}}=\lambda_{\text {NEIES}} \times N_{\text {target}},
\end{aligned}
\end{equation}
where $N_{\text{target}}$ is the number of target particles.

\section{RESULTS AND DISCUSSION}
\label{section 3}
\begin{figure*}
\centering	
\includegraphics[width=17.5cm,height=9cm]{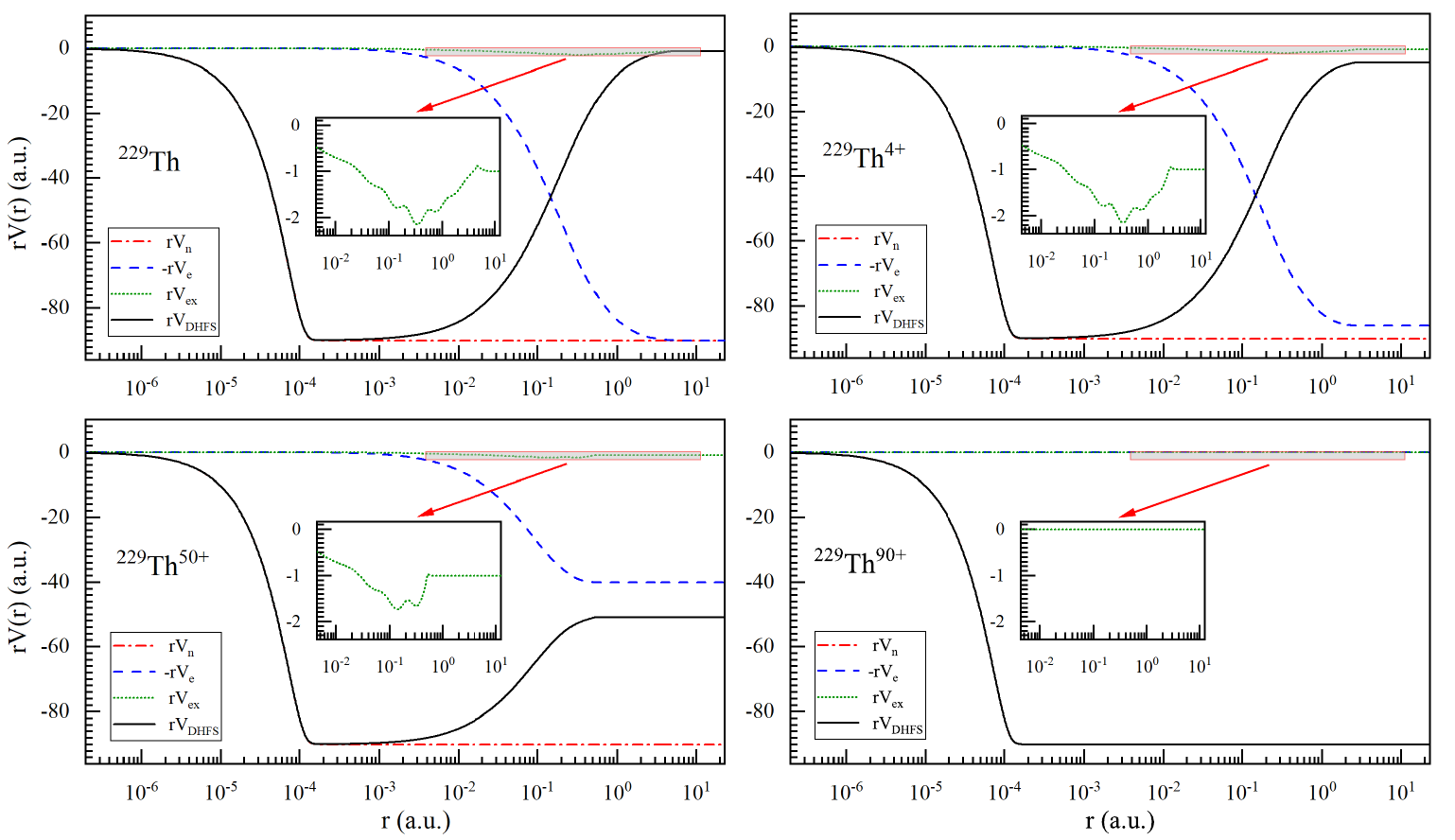} 
\caption{DHFS self-consistent potential $r V_{\text {DHFS}}(r)$ for Th and selected Th ions ($\mathrm{Th}^{4+}$, $\mathrm{Th}^{50+}$, $\mathrm{Th}^{90+}$). The nuclear, electronic, exchange, and total potentials are represented by red dot-dashed, blue dashed, green short-dotted, and black solid lines, respectively.}
\label{fig 1}
\end{figure*}
In this study, we focus on the ${ }^{229} \mathrm{Th}$ atom and its ions in various charge states, using the DHFS method to calculate potential energy distributions. We present the potential energy curves for neutral $\mathrm{Th}$ atom, $\mathrm{Th}^{4+}$, $\mathrm{Th}^{50+}$ and $\mathrm{Th}^{90+}$ ions in Fig. \ref{fig 1}. These charge states help us understand the impact of ionization on the radial wavefunctions, which are crucial for determining the excitation cross sections.

From Fig. \ref{fig 1}, it can be observed that the nuclear potential remains largely unchanged across all ionization states, as it only depends on the proton charge and nuclear size. The exchange potential shows minimal variation across ionization states. To better show these changes, the exchange potential for each state is magnified and displayed separately in the figure. By combining the main plot and the magnified views, we observe that both the exchange potential and the electronic potential decrease with increasing ionization, becoming negligible for the highly ionized $\mathrm{Th}^{90+}$ state due to the absence of bound electrons. The total potential varies significantly with different ionization states because of changes in the electronic potential. For neutral and low-ionization states, the electron cloud extends far from the nucleus, causing a gradual decay in total potential at larger distances due to significant electron shielding. In contrast, for highly ionized states, most electrons are stripped away, resulting in minimal electron shielding. Thus, the total potential, dominated by the nuclear Coulomb potential in highly ionized states, shows sharp changes near the nucleus and diminishes rapidly at larger distances.

To further explore the excitation mechanisms of ${}^{229} \mathrm{Th}$, we examine the nuclear states and their properties. For the ${}^{229}\mathrm{Th}$ nucleus, the spin-parities of the ground state (GS), IS and SE are $5/2^{+}$, $3/2^{+}$ and $5/2^{+}$, respectively. Following the selection rules for multipole radiation and parity conservation, transitions from the GS to both the IS and SE can occur via magnetic dipole (M1) or electric quadrupole (E2) modes. Theoretical calculations give the reduced transition probabilities: for the IS to GS transition, $B(E2) = 27$ W.u. and $B(M1) = 0.0076$ W.u., while for the SE to GS transition, $B(E2) = 39.49$ W.u. and $B(M1) = 0.0043$ W.u. \cite{PhysRevLett.118.212501,PhysRevLett.122.162502,PhysRevC.103.014313,PhysRevC.105.064313}.
\begin{figure}[H]
	\includegraphics[width=9.0cm,height=8.5cm]{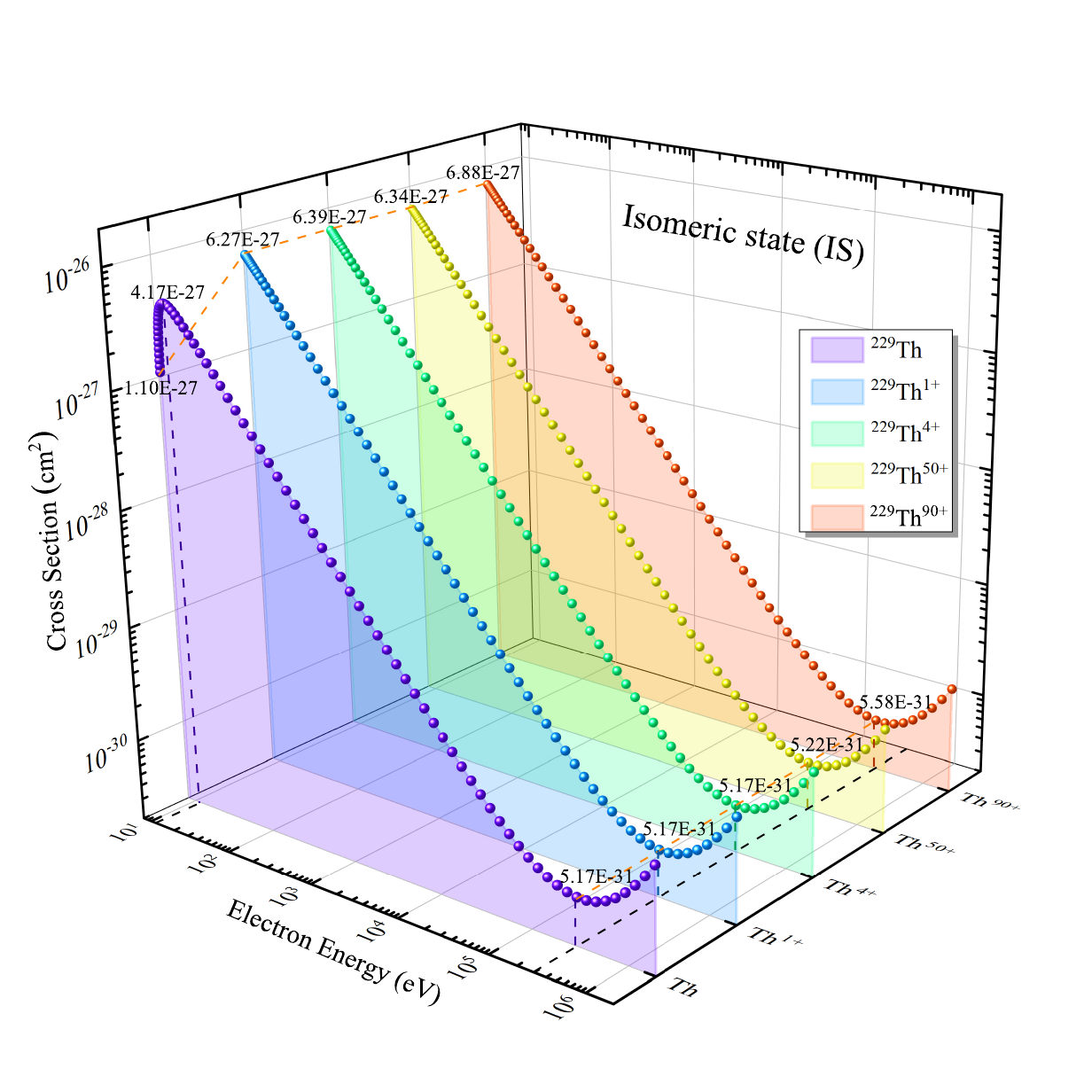}
	\caption{NEIES cross sections for exciting the isomeric state (IS) of Th atom, Th$^{1+}$, Th$^{4+}$, Th$^{50+}$ and Th$^{90+}$.}
	\label{fig 2}
\end{figure}
We calculate the excitation cross sections of ${}^{229}\mathrm{Th}$ over a broad range of incident electron energies: 8.4 eV to 2 MeV for IS and 29.19 keV to 2 MeV for SE. Although electron energies up to 2 MeV could, in principle, excite the nucleus to higher energy states beyond the SE, we focus on the IS and SE transitions. This is because excitation cross sections for higher states are significantly smaller, and their complex decay pathways contribute negligibly to the accumulation of nuclei in the IS \cite{PhysRevLett.125.142503}. These calculations are carried out for the neutral $\mathrm{Th}$ atom and several ionized states—$\mathrm{Th}^{1+}$, $\mathrm{Th}^{4+}$,  $\mathrm{Th}^{50+}$ and $\mathrm{Th}^{90+}$. The results for IS and SE are shown in Fig. \ref{fig 2} and Fig. \ref{fig 3}.

In Fig. \ref{fig 2}, the excitation cross section $\sigma_{1}$ for the neutral ${}^{229}\mathrm{Th}$ atom increases from $1.10 \times 10^{-27}$ cm$^2$ at 8.4 eV to a peak of $4.17 \times 10^{-27}$ cm$^2$ at 10.8 eV. This increase is due to the dense electron cloud around the nucleus, which shields the nucleus and reduces effective electron-nucleus interactions. As the electron energy exceeds the excitation threshold, higher kinetic energy allows electrons to better penetrate the shielding, enhancing overlap with the nuclear region and thus increasing the excitation probability. As energy increases, the cross section decreases sharply, reaching a minimum of $5.17 \times 10^{-31}$ cm$^2$ around 300 keV. This decline results from the spreading of higher-energy electron wavefunctions, which reduces effective nuclear interactions. Additionally, enhanced electron shielding further diminishes the cross section. Beyond 300 keV, the cross section rises again. For ionized ${}^{229}\mathrm{Th}$ species, the excitation cross section peaks at 8.4 eV, unlike the gradual rise in the neutral atom. This difference arises because reduced electron shielding in ionized states permits electrons to interact directly with the nucleus once the excitation threshold is reached. After the peak, the cross section for ionized states follows a similar trend to that of the neutral atom. Fig. \ref{fig 3} shows a similar trend for the SE, with excitation cross sections $\sigma_{2}$ beginning at a higher threshold energy of 29.19 keV. Initially, the cross sections decrease, reaching a minimum around 219 keV, then steadily increase as electron energy rises. This behavior mirrors that of the IS but occurs at higher energy levels because of the greater excitation energy requirement.
\begin{figure}[H]
\includegraphics[width=9.0cm,height=8.5cm]{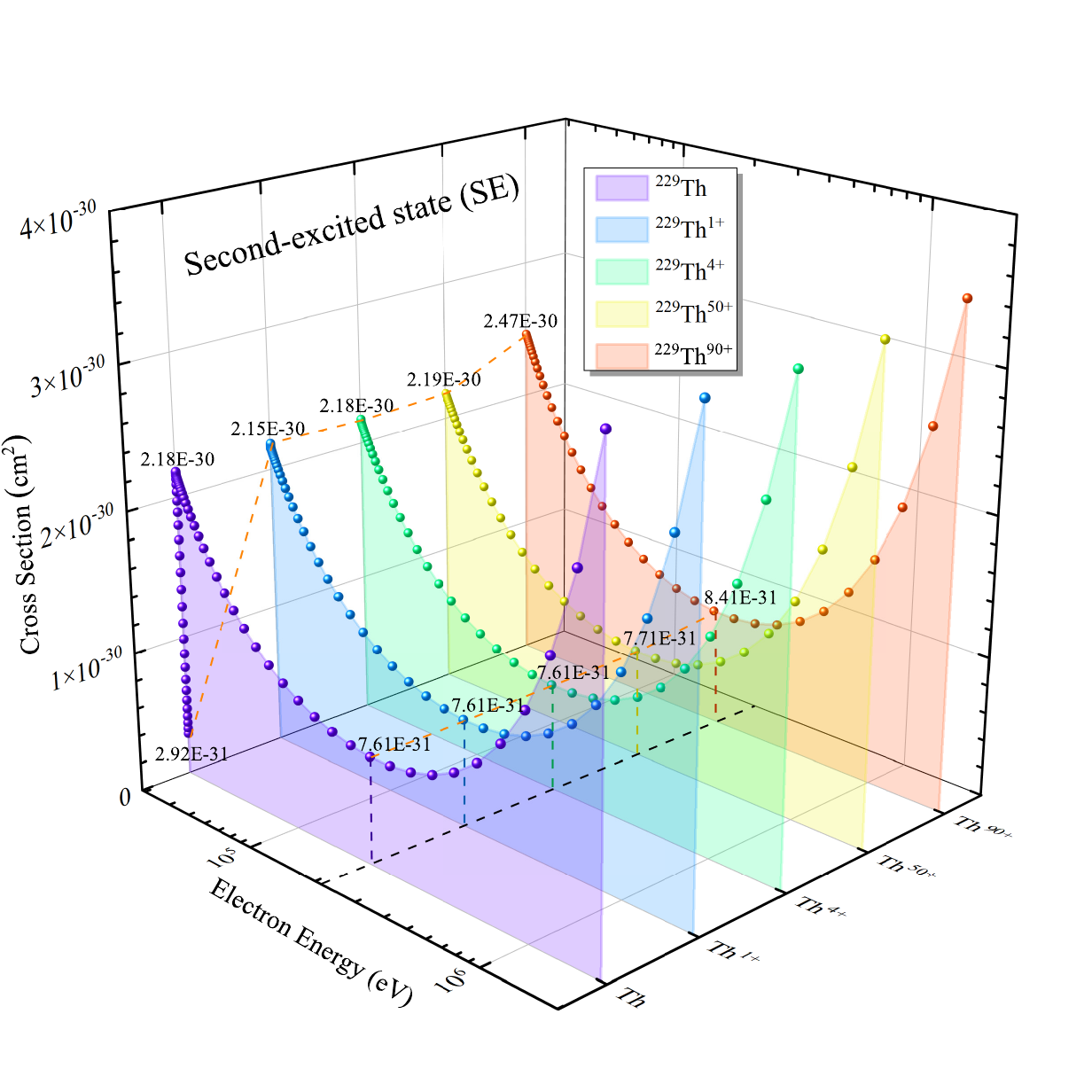}
\caption{NEIES cross sections for exciting the second-excited state (SE) of Th atom, Th$^{1+}$, Th$^{4+}$, Th$^{50+}$ and Th$^{90+}$.}
\label{fig 3}
\end{figure}
In general, increased ionization generally leads to larger cross sections, but the differences remain within the same order of magnitude. While higher ionization reduces electron shielding and enhances electron-nucleus interactions, its overall impact on the cross section is moderate, with remaining electron density and relativistic effects also playing a role.

\begin{figure}[H]
	\includegraphics[width=8.8cm,height=5.8cm]{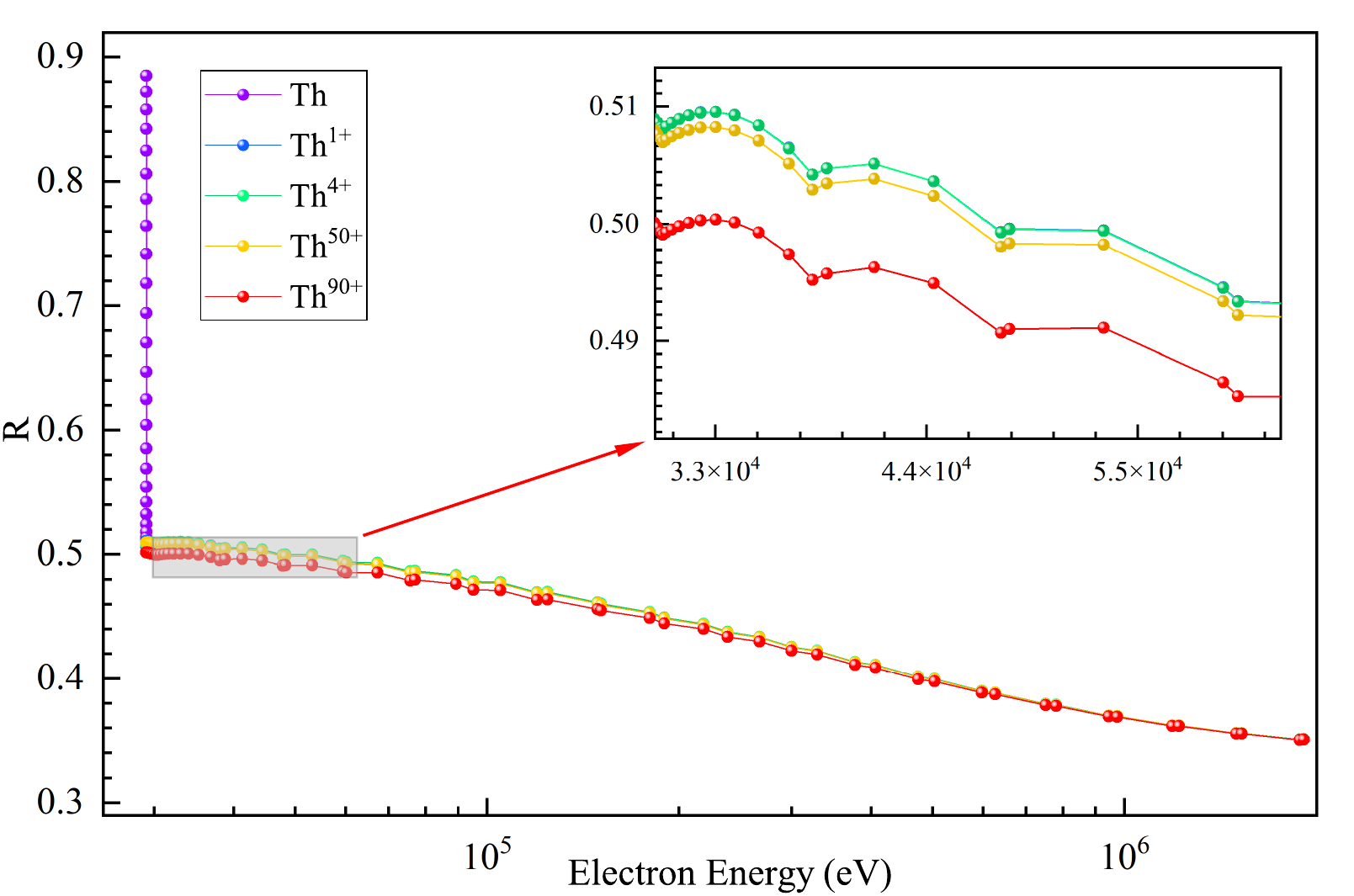}
	\caption{Excitation ratio $R$ as a function of incident electron energy $E$ for Th atom, $\mathrm{Th}^{1+}$, $\mathrm{Th}^{4+}$, $\mathrm{Th}^{50+}$ and $\mathrm{Th}^{90+}$.}
	\label{fig 4}
\end{figure}

Provided that the incident electron energy exceeds 29.19 keV, electrons can excite the  ${}^{229} \mathrm{Th}$ nucleus to the IS and SE, with the probabilities depending on. We define the excitation ratio $R$ as
\begin{equation}
\label{eq16}
\begin{aligned}
R=\frac{\sigma_1}{\sigma_1+\sigma_2},
\end{aligned}
\end{equation}
where $\sigma_1$ and $\sigma_2$ are the cross sections for excitation to the IS and SE, respectively. The proportion leading to the SE is $(1-R)$. We calculated $R$ for various ionization states of ${}^{229}\mathrm{Th}$ and plotted the results in Fig. \ref{fig 4}. To better illustrate the differences, we magnified the region where electron energies are between $30$ $\mathrm{keV}$ and $60$ $\mathrm{keV}$.
In Fig. \ref{fig 4}, when the electron energy exceeds 29.19 keV, the excitation ratio $R$ for the neutral ${}^{229}\mathrm{Th}$ approaches 1, indicating that nearly all excitations lead to the first excited state. As the energy increases, $R$ decreases rapidly to around 0.5. In contrast, for ionized states, $R$ starts near 0.5 across the entire energy range. As the energy continues to rise, $R$ gradually decreases towards 0.34 for all ionization states, mirroring the trend seen in the neutral atom. The magnified section of Fig. \ref{fig 4} reveals significant differences in $R$ among various ionization states near 30 keV, but these differences diminish as the energy increases, indicating that at higher energies, the influence of ionization on the excitation pathway becomes negligible. The trend in $R$ reflects the role of electron binding energies and kinetic energy in nuclear excitation. In neutral atoms, lower binding energies favor initial excitation to the IS, while higher energies increase the probability of SE excitation, reducing $R$. In ionized states, tighter binding initially balances excitation probabilities, but the trend eventually mirrors that of the neutral state at high energies.

To accurately determine the effective excitation rate $\lambda_{\mathrm{eff}}$, we integrate the excitation cross sections with the electron flux distribution $\Phi_e(E)$. This is expressed as
\begin{equation}
\label{eq17}
\begin{aligned}
\lambda_{\text {eff }}=\lambda_{1}+\mathrm{BR} \times \lambda_{2},
\end{aligned}
\end{equation}
where $\lambda_1$ and $\lambda_2$ represent the excitation rates directly to the IS and SE, respectively, and $\mathrm{BR}$ is the branching ratio (approximately 90\%) from SE to IS \cite{PhysRevC.110.014330}. Based on the electron beam parameters, the electron flux $\Phi_e(E)$ is modeled as a Gaussian distribution, so we calculate $\lambda_1$ and $\lambda_2$ via numerical integration
\begin{equation}
\label{eq18}
\begin{aligned}
\lambda_{1,2}=\frac{I}{e A_{\text{beam}} \epsilon_e \sqrt{\pi}} \int \sigma_{1,2}(E) \exp \left[-\left(\frac{E-E_r}{\epsilon_e}\right)^2\right] d E.
\end{aligned}
\end{equation}
Here, $I$ is the electron beam current (200 mA), $e$ is the elementary charge ($1.602 \times 10^{-19}$ C), and $A_{\text{beam}}$ is the beam cross-sectional area, determined from a radius of $50$ $\mu$m \cite{10.3389/fphy.2023.1203401}.
\begin{figure}[H]
	\includegraphics[width=9.0cm]{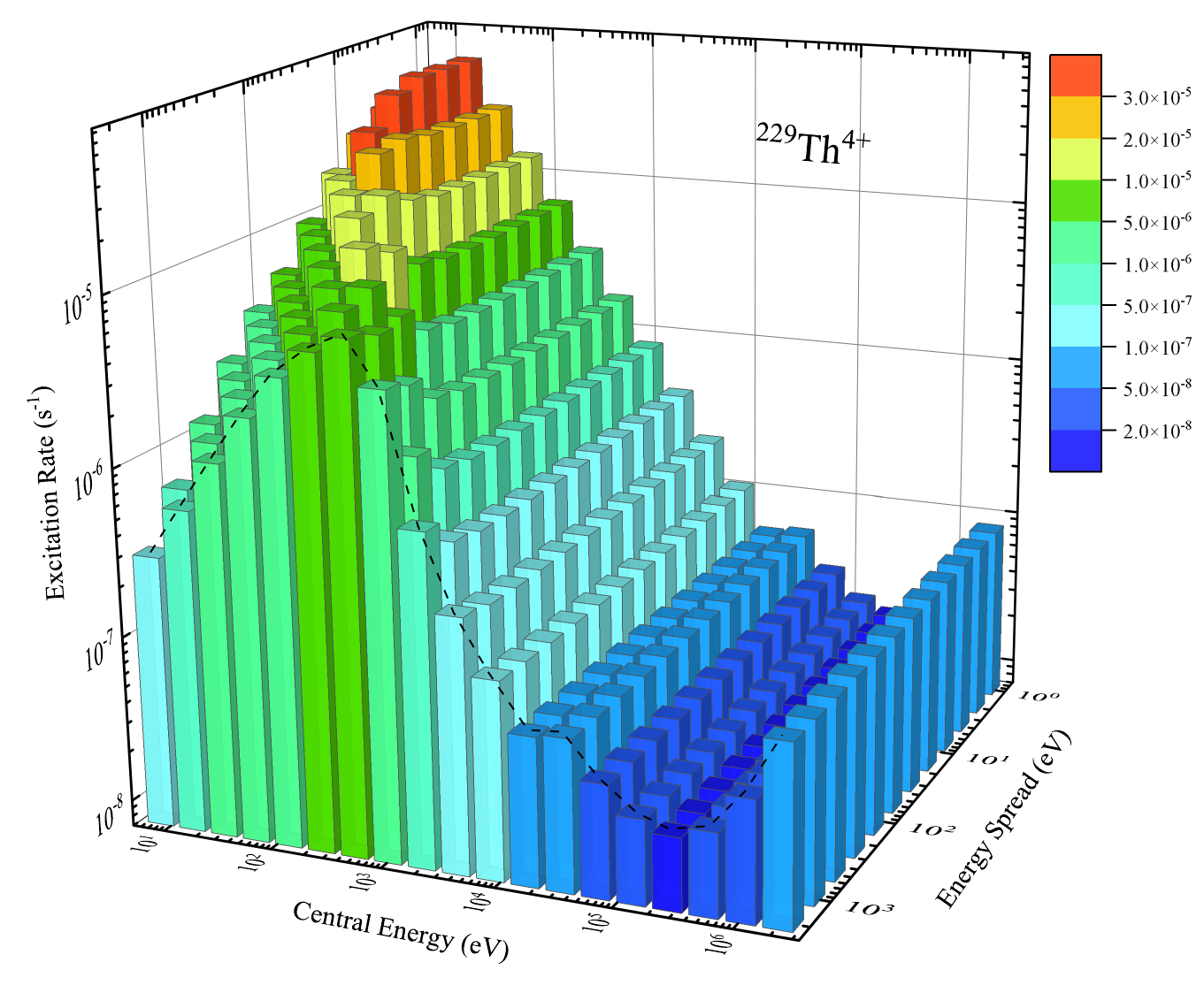}
	\caption{Three-dimensional visualization of the effective excitation rate $\lambda_{\mathrm{eff}}$ for Th$^{4+}$ as a function of central energy $E_r$ and energy spread $\epsilon_e$.}
	\label{fig 5}
\end{figure}
Fig. \ref{fig 5} presents a three-dimensional visualization of the effective excitation rate $\lambda_{\mathrm{eff}}$ for Th$^{4+}$ as a function of central electron energy $E_r$ and energy spread $\epsilon_e$, assuming a Gaussian electron energy distribution. The data indicates a clear dependence of $\lambda_{\mathrm{eff}}$ on both parameters. Specifically, when $\epsilon_e$ is held constant, increasing $E_r$ results in a non-monotonic behavior of $\lambda_{\mathrm{eff}}$. As $E_r$ approaches the excitation threshold, $\lambda_{\mathrm{eff}}$ initially rises, primarily because the central energy being close to the excitation energy allows some electrons to fall below the effective excitation region. With a larger energy spread $\epsilon_e$, this rise becomes more pronounced. However, beyond the optimal region near the excitation threshold, $\lambda_{\mathrm{eff}}$ first declines and then rises again, reflecting the behavior of the excitation cross-section. An intriguing observation is the unexpected increase in $\lambda_{\mathrm{eff}}$ within the high-energy range of 30 keV to 50 keV, which contrasts with the anticipated continuous decline. This increase can be attributed to the involvement of the SE. Given that approximately 90\% of SE decays into IS, $\lambda_{\mathrm{eff}}$ benefits from both direct excitation of IS and indirect excitation via SE, leading to a higher overall rate. The diminishing impact of energy spread $\epsilon_e$ on excitation rates at higher $E_r$ suggests that once electron energy exceeds the excitation thresholds, precise energy distribution tuning is less crucial. This flexibility is advantageous for experimental setups operating in the high-energy regime, where maintaining a narrow energy spread is technically challenging.

In addition to general trends, Fig. \ref{fig 5} also highlights specific regions of high or low excitation efficiency. In the low-energy region, the highest excitation rate occurs at $E_r = 8$ eV with $\epsilon_e = 1$ eV, resulting in $\lambda_{\mathrm{eff}}$ reaching $3.24 \times 10^{-5}$ s$^{-1}$. This configuration ensures that most electrons have optimal energies for nuclear excitation, thereby maximizing the probability of exciting $^{229}$Th to its IS. Experiments can target these parameters to enhance the production efficiency of the IS. Conversely, certain parameter sets lead to significantly reduced excitation rates. For instance, at $E_r = 262$ keV, $\lambda_{\mathrm{eff}}$ drops to $1.93 \times 10^{-8}$ s$^{-1}$, indicating a low-excitation region. Such low-rate regions are particularly important for studying NEEC \cite{PhysRevLett.130.112501}, which has not yet been experimentally observed. Conducting experiments in these regions helps reduce interference from competing processes like NEIES, thereby creating a cleaner experimental environment more favorable for detecting NEEC.

\begin{figure*}
	\centering	
	\includegraphics[width=17.5cm]{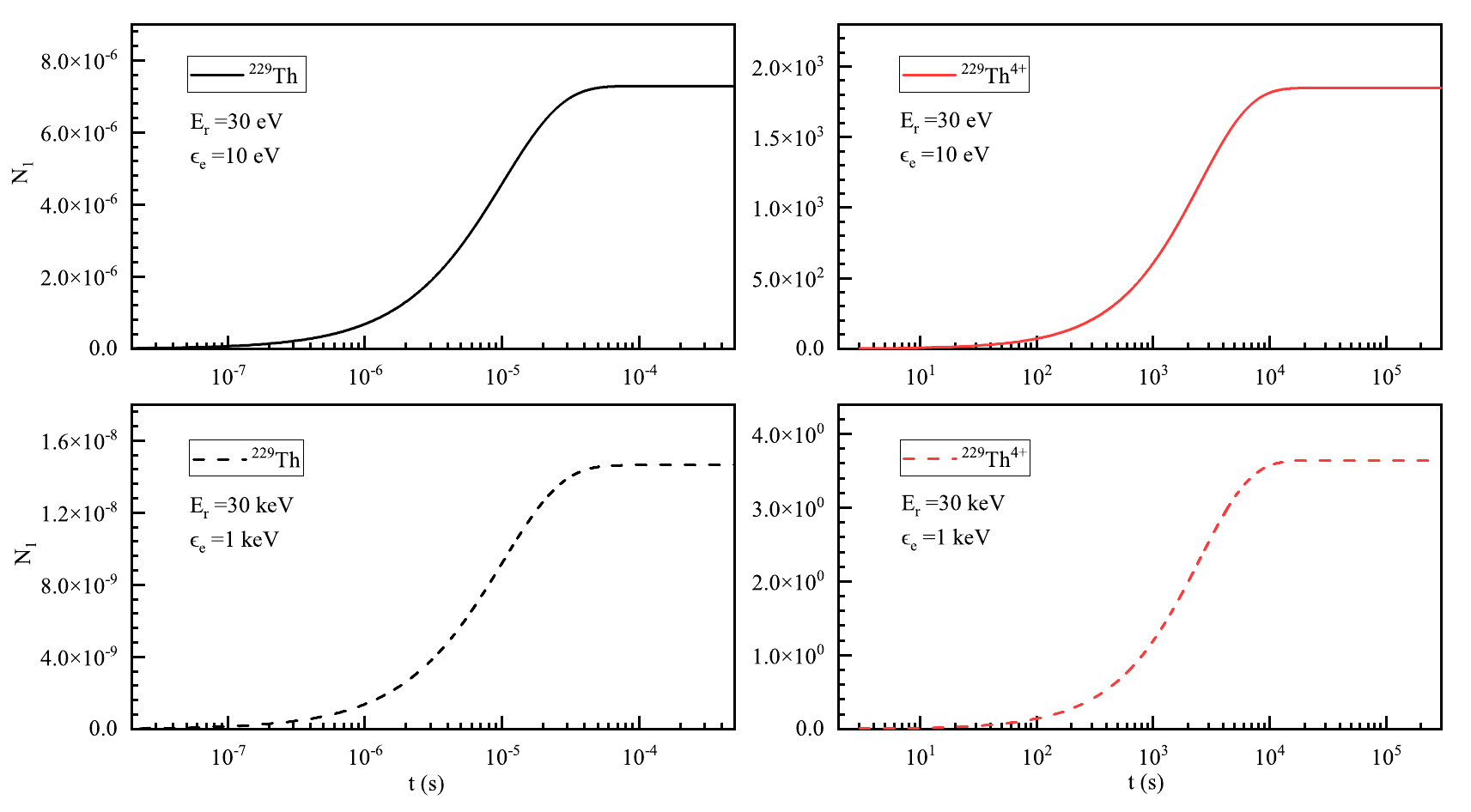}
	\caption{Temporal evolution of the isomeric state population $N_1(t)$ for neutral ${}^{229}\mathrm{Th}$ atom and $\mathrm{Th}^{4+}$ ion under low-energy ($E_r=30$ eV) and high-energy ($E_r=30$ keV) excitation.}
	\label{fig 6}
\end{figure*}
To understand the temporal evolution of the IS population in the $^{229}$Th nucleus, we develop a dynamic model that describes both excitation and decay processes. We model the populations of IS ($N_1(t)$) and SE ($N_2(t)$), represented by
\begin{equation}
\label{eq19}
\begin{aligned}
\frac{d N_1(t)}{d t}= \lambda_1 N_{\text {target }} + \mathrm{BR} \times\Gamma_{2 \rightarrow 1} N_2(t) - \Gamma_1 N_1(t)
\end{aligned}
\end{equation}
and
\begin{equation}
\label{eq20}
\begin{aligned}
\frac{d N_2(t)}{d t}= \lambda_2 N_{\text {target }} - \Gamma_{2 \rightarrow 1} N_2(t).
\end{aligned}
\end{equation}
Here, $N_{\text{target}}$ represents the total number of target particles. Assuming a trap with the radius of 50 $\mu\text{m}$, the length of 3 cm, and the target density of $10^8$ $\text{cm}^{-3}$ results in $N_{\text{target}} \approx 2.36 \times 10^4$ \cite{10.3389/fphy.2023.1203401}. $\Gamma_1$ and $\Gamma_{2 \rightarrow 1}$ are the decay rates of the IS and SE, respectively, calculated using $\Gamma = \ln 2 / T_{1/2}$, where $T_{1/2}$ is the half-life.

We select neutral $^{229}$Th atom and Th$^{4+}$ ion for comparison because of their distinct decay characteristics. For neutral $^{229}$Th, the IS has a short half-life of about 7 microseconds, with a decay rate of $\Gamma_1 \approx 9.90 \times 10^4$ $\text{s}^{-1}$ \cite{PhysRevLett.118.042501}. The SE in neutral $^{229}$Th has a half-life of 82.2 picoseconds, with $\Gamma_{2 \rightarrow 1} \approx 8.44 \times 10^9$ $\text{s}^{-1}$ \cite{masuda2019x}. In contrast, for Th$^{4+}$ ions, internal conversion is largely suppressed, extending the half-life of the IS to 1740 seconds, corresponding to $\Gamma_1 \approx 3.98 \times 10^{-4}$ $\text{s}^{-1}$ \cite{PhysRevLett.132.182501}. The SE decay rate remains unchanged.

Given $\Gamma_{2 \rightarrow 1} \gg \Gamma_1$, we use the quasi-steady-state approximation for the SE, and we get
\begin{equation}
\label{eq21}
\begin{aligned}
N_2(t) \approx \frac{\lambda_2 N_{\text {target }}}{\Gamma_{2 \rightarrow 1}}.
\end{aligned}
\end{equation}
Substituting this into the rate equation for $N_1(t)$
\begin{equation}
\label{eq22}
\begin{aligned}
\frac{d N_1(t)}{d t}=\left(\lambda_1+\mathrm{BR} \times \lambda_2\right) N_{\text {target }}-\Gamma_1 N_1(t).
\end{aligned}
\end{equation}
Solving this first-order linear differential equation with the initial condition $N_1(0) = 0$ yields
\begin{equation}
\label{eq23}
\begin{aligned}
N_1(t)=\frac{\left(\lambda_1+\mathrm{BR}\times\lambda_2\right)N_{\text{target}}}{\Gamma_1}\left(1-e^{-\Gamma_1 t}\right).
\end{aligned}
\end{equation}
At steady state ($t \to \infty$), the population of the IS reaches
\begin{equation}
\label{eq24}
\begin{aligned}
N_1(\infty)=\frac{\left(\lambda_1+\mathrm{BR} \times \lambda_2\right) N_{\text {target }}}{\Gamma_1}.
\end{aligned}
\end{equation}

In Fig. \ref{fig 6}, we illustrate the temporal evolution of the IS population $N_1(t)$ for both neutral ${}^{229}\mathrm{Th}$ atoms and $\mathrm{Th}^{4+}$ ions under two excitation regimes: low-energy ($E_r = 30$ eV, $\epsilon_e = 10$ eV) and high-energy ($E_r = 30$ keV, $\epsilon_e = 1$ keV). The black lines represent neutral atoms, while the red lines represent Th$^{4+}$ ions. Solid lines correspond to the low-energy regime, primarily exciting the IS, while dashed lines show the high-energy regime, exciting both the IS and SE ($\lambda_2$ becomes non-negligible).

The four subplots in Fig. \ref{fig 6} reveal similar trends but with notable numerical differences. Under low-energy excitation, $N_1$ in neutral atoms rises rapidly, stabilizing at approximately $7.29 \times 10^{-6}$ within tens of microseconds, while $\mathrm{Th}^{4+}$ ions exhibit a much slower rise, taking several thousand seconds to reach a steady-state value of around $1.85 \times 10^3$. This difference is due to their different half-lives and the longer half-life of $\mathrm{Th}^{4+}$ allows for gradual accumulation, leading to a much higher IS population compared to neutral atoms. Notably, this steady-state value for $\mathrm{Th}^{4+}$ ions represents nearly 10\% of the total target number ($N_{\text{target}} \approx 2.36 \times 10^4$), demonstrating an efficient excitation process. Such efficiency underscores the potential for using low-energy electron beams in applications like nuclear clocks and advanced spectroscopy, where maintaining a substantial population in the desired state is crucial. In the high-energy regime, both the IS and SE are excited, but the overall efficiency is lower than in the low-energy regime. For neutral atoms, $N_1$ stabilizes at approximately $1.46 \times 10^{-8}$, while for $\mathrm{Th}^{4+}$ ions, it reaches a steady-state value of $3.64$. This reduction is mainly caused by the lower excitation cross-sections associated with the higher energy states, resulting in less effective population accumulation.

\section{Summary}
In this study, we systematically investigate the excitation dynamics of $^{229}$Th nuclei through inelastic electron scattering, focusing on the effects of varying electron energy and ionic charge states. Using the DHFS method, we analyze the differences in potential energy experienced by electrons in different ionization states. We calculate the excitation cross sections across multiple ionization states, showing that higher degrees of ionization moderately enhance the excitation cross sections. In the high-energy region (above 29.19 keV), electrons can excite both the IS and SE, which in turn boosts the overall excitation efficiency of the isomeric state. By applying rate equations, we model the accumulation of the IS population over time and find that under certain conditions, up to 10\% of the Th$^{4+}$ nuclei can be accumulated in the IS. This research highlights how adjusting electron energy and ionization levels can optimize the excitation process of $^{229}$Th nuclei, offering valuable theoretical support for applications such as nuclear clocks and precision measurements.

\label{section 4}

\begin{acknowledgments}
This work was supported by the National Natural Science Foundation of China (Grant No.12135009, 12375244), the Hunan Provincial Innovation Foundation for Postgraduate (Grant No.CX20230008), and the Innovation Foundation for Postgraduate (Grant No.XJQY2024046, XJJC2024063).
\end{acknowledgments}

\bibliographystyle{apsrev4-1}

\end{document}